\documentstyle[twocolumn,aps,epsfig]{revtex} 
 
\draft 
\begin{document} 

 
\title{$\mu$SR measurement of the fundamental length scales in the vortex
state of YBa$_2$Cu$_3$O$_{6.60}$} 
 
\author{ J.~E.~Sonier, J.~H.~Brewer, R.~F.~Kiefl\\ 
D.~A.~Bonn, S.~R.~Dunsiger, W.~N.~Hardy, Ruixing~Liang, W.~A.~MacFarlane,
R.~I.~Miller and T.~M.~Riseman 
\thanks{Present address: Superconductivity Research Group, 
University of Birmingham, 
Birmingham, B15 2TT, United Kingdom}} 

\address{TRIUMF, Canadian Institute for Advanced Research 
 and Department of Physics and Astronomy, University of British Columbia, 
 Vancouver, British Columbia, Canada V6T 1Z1 } 

\author{ D.~R.~Noakes, C.~E.~Stronach and M.~F.~White~Jr.} 
 
\address{Department of Physics, Virginia State University, 
Petersburg, Virginia 23806}

\date{June, 1997} 
\date{ \rule{2.5in}{0pt} } 
 
\maketitle 
\begin{abstract} \noindent 

The internal field distribution in the vortex state of
YBa$_2$Cu$_3$O$_{6.60}$ is shown to be a sensitive measure of both
the magnetic penetration depth $\lambda_{ab}$ and the vortex-core
radius $\rho_0$. The temperature dependence of $\rho_0$ is 
found to be weaker than in the conventional superconductor NbSe$_2$
and much weaker than theoretical predictions for an isolated
vortex. The effective vortex-core radius decreases sharply 
with increasing $H$, whereas 
$\lambda_{ab}(H)$ is found to be much stronger than in NbSe$_2$. 
\end{abstract} 
\begin{center}
\rule{0.3\textwidth}{0.25pt} 
\end{center}
 
The magnetic penetration depth $\lambda$ and the coherence
length $\xi$ are the two fundamental length scales in a
superconductor. $\lambda$ is directly related to the superfluid density $n_s$ 
whereas $\xi$ is the length scale for spatial variations
in the superconducting order parameter.
In recent years, measurements of $\lambda$ have provided 
strong evidence for unconventional pairing in the high-$T_c$ materials
in that there are line nodes in the superconducting energy gap. 
In particular, in both the Meissner and vortex states, 
$\lambda$ has been observed
to increase linearly with temperature $T$ \cite{Hardy:93,Sonier:94}
and magnetic field $H$ \cite{Maeda:96,Sonier:97} at low $T$ in single
crystals of YBa$_2$Cu$_3$O$_{6.95}$.

Much less is known about the behaviour of $\xi$. In fact
until now there has been no measurement of $\xi$ deep in the
superconducting state of a high-$T_c$ superconductor.
The magnitude of $\xi$ near the phase boundary has
been estimated from measurements of the upper critical field $H_{c2}$
using Ginzburg-Landau (GL) theory.
For a type-II superconductor for $H$ near $H_{c2}$, the coherence length
is related to $H_{c2}(T)$ in GL theory by,
\begin{equation}
\xi (T) = \sqrt{\frac{\Phi_{\circ}}{2 \pi H_{c2}(T)}}\,\,.
\label{eq:Coherence} 
\end{equation}
Reliable estimates of $\xi$ in YBa$_{2}$Cu$_{3}$O$_{7-\delta}$
made in this way are extremely difficult because the value of $H_{c2}$
at $T\!=\!0$ is so large (i.e. $\sim \! 100$~T). Consequently, $H_{c2}$
measurements are limited to high temperatures above $T/T_c \sim 0.85$, 
so that the low $T$ behaviour
of $\xi (T)$ can only be determined by extrapolation.
The situation is further complicated in the high-$T_c$ compounds
by strong thermal fluctuations over a sizeable region near $T_c$
which result in broad transitions
and poor estimates of $H_{c2}(T)$.
Furthermore, Eq.~(\ref{eq:Coherence}) may not be valid for an unconventional 
superconductor.    

It is desireable to have direct measurements of the coherence length,
which in the vortex state is related to the size of the vortex cores.
In particular, for a conventional superconductor $\xi \! \sim \! \rho_0$,
where $\rho_0$ is the vortex-core radius \cite{Caroli:64}.
In principle both STM and $\mu$SR can be used to characterize the
size of vortex cores and thereby determine $\xi$ directly deep
in the superconducting state.
Recent STM \cite{Hartmann:93} and muon spin rotation ($\mu$SR) \cite{nbse2:97} 
measurements on NbSe$_2$ show that $\rho_0$
decreases with increasing $H$, as a result of 
the increased interaction between vortices.
Similar studies on high-$T_c$ materials have not yet been performed with
STM. Pinning of the vortices
due to surface roughness and oxygen vacancies eliminates the long-range 
order in the vortex lattice and results in variations in the 
electronic structure of the cores. 
Furthermore, the effect of the discontinuity in the quasiparticle
excitation spectrum at the surface is still not understood.

On the other hand, $\mu$SR provides information
on the vortex cores in the bulk of the sample.
As explained in Ref.~\cite{nbse2:97}, in a $\mu$SR experiment
$\rho_0$ is related to 
the high-field cutoff of the measured internal field 
distribution. The spectral weight at the cutoff grows as the density 
of the vortices increases.
A well defined cutoff was observed in
NbSe$_2$ (Ref.~\cite{nbse2:97}), where
$\rho_0$ is several times larger than in YBa$_2$Cu$_3$O$_{6.95}$ 
(Refs.~\cite{Sonier:94,Sonier:97}). In YBa$_2$Cu$_3$O$_{6.95}$ no clear
signal from the vortex cores was visible below 3~T.
The temperature dependence of $\xi (T)$ was investigated in 
an earlier $\mu$SR study of YBa$_2$Cu$_3$O$_{6.95}$ at higher
magnetic fields \cite{Riseman:95}. Unfortunately 
the signal-to-noise ratio was poor due to the
influence of the large magnetic field on the positron orbits
and timing resolution.
In addition, demagnetization effects likely contributed significantly
to the measured field distribution since the sample consisted
of nineteen small crystals, and $\lambda$ and $\xi$ were
assumed independent of $H$ in the fitting procedure.

In this Letter we present a $\mu$SR study of 
the oxygen deficient high-$T_c$ superconductor YBa$_2$Cu$_3$O$_{6.60}$. 
Compared to the optimally-oxygenated compound, YBa$_2$Cu$_3$O$_{6.60}$
has a smaller carrier concentration in the CuO$_2$ planes, a reduced
$T_c$ and $H_{c2}$, and a correspondingly larger $\xi$.
This allows us now to report the first detailed study of
the fundamental length scales $\lambda$ and $\xi$ in a high-$T_c$
superconductor.
  
We studied two different YBa$_{2}$Cu$_{3}$O$_{6.60}$ samples with identical 
transition temperatures (59~K). The first sample (S1)
was obtained by deoxygenating the three-crystal mosaic of
YBa$_{2}$Cu$_{3}$O$_{6.95}$ used in Refs.~\cite{Sonier:94,Sonier:97}.
The twin boundary spacing was on the order of $1~\mu$m in the bulk.
The second  sample (S2) was grown from a separate batch and 
consisted of two large 
single crystals with a total $\hat{a}$-%
$\hat{b}$ plane surface area of 30~mm$^2$.
S2 was mechanically detwinned, such that the
twin boundary density was about
an order of magnitude smaller than in S1. 
Measurements were performed on field-cooled samples using the 
M15 and M20 surface muon beamlines at TRIUMF and
the same apparatus as that used in Refs.~\cite{Sonier:94,Sonier:97,nbse2:97}.
 
The experimental muon spin precession signal
was fit assuming the local field due to the vortex lattice
at any point in the $\hat{a}$-%
$\hat{b}$ is given by \cite{Yaouanc:97}, 
\begin{eqnarray} 
\label{eq:Br} 
 B(\mbox{\boldmath $\rho$}) & = & B_0 (1-b^4)\sum_{ {\bf G} }
 { e^{-i {\bf G} \cdot \mbox{\boldmath $\rho$} }
 \,\, u \, K_1(u)
 \over
 \lambda_{ab}^2 G^2}, \eqnum{2a} \\
\nonumber \\ 
\mbox{with,} \;\;\;\;\;
u^2 & = & 2 \, \xi_{ab}^2 G^2 (1+b^4)[1-2b(1-b)^2]. \eqnum{2b}
\end{eqnarray} 
where $B_0$ is the average magnetic field, $b=B/B_{c2}$,
$\xi_{ab}$ is the GL coherence length and $K_1(u)$ is a
modified Bessel function. This analytical model of the field profile
agrees extremely well with the exact numerical solutions of
the GL equations \cite{Brandt:97} at low reduced fields $b$---whereas
$b<0.02$ for our measurements. 
The term $u \, K_1(u)$ cuts off the summation thereby
removing the logarithmic divergence of field at the vortex cores
in the conventional London model. The cutoff is done in a way which preserves
circular symmetry around the vortex cores.
We note that the results herein are qualitatively similar to that obtained
using a modified London model for $B(\rho)$ with a Gaussian cutoff factor
(as in Ref.~\cite{Sonier:97}) and nearly identical to that using a Lorentzian
cutoff---where only the latter is strictly valid at low fields. 
The internal field distribution
$n(B)$ was convoluted with a Gaussian of width $\sigma$ to
account for vortex-lattice disorder and nuclear dipolar fields.

The summation in Eq.~(\ref{eq:Br}) is taken over all
reciprocal lattice vectors {\bf G} of a triangular vortex 
lattice---the structure which minimizes the free energy for a conventional
superconductor. Infrared reflectance measurements 
of $\lambda$ in zero-field \cite{Basov} show
a small anisotropy in the $\hat{a}$-%
$\hat{b}$ plane ($\lambda_a/\lambda_b \! \cong \! 1.3$)
which will stretch
the triangular lattice, leading to elliptical cores
in which $\xi_b/\xi_a \! = \! \lambda_a/\lambda_b$. 
It was shown in Ref.~\cite{Sonier:97} through a scaling
argument, that the corresponding magnetic field distribution is identical 
to the isotropic case.
\begin{figure}[t]
\epsfig{figure=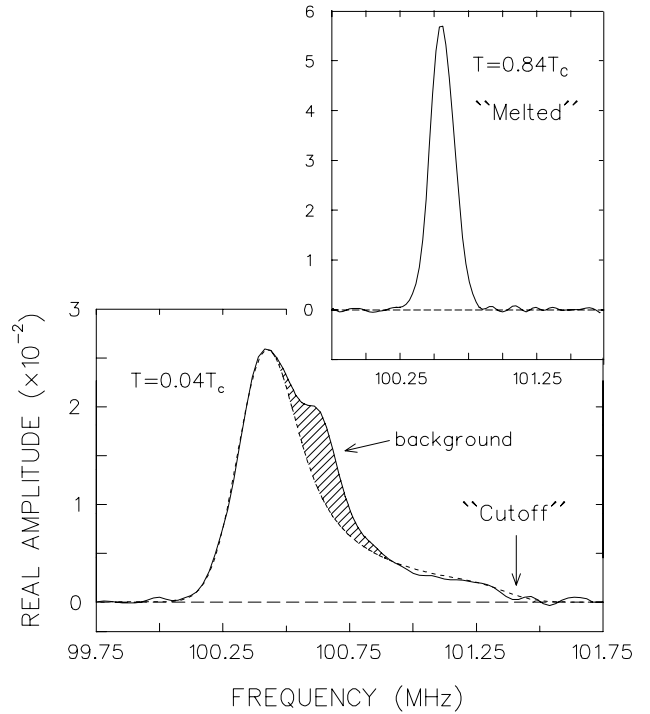, width=\linewidth}
\\
\caption[]{
The Fourier transforms of the muon spin precession
signal in YBa$_2$Cu$_3$O$_{6.60}$ after field cooling in a magnetic
field $H \sim$0.75~T down to $T \! = \! 0.04$ and
$0.84~T_c$ (inset). The dashed curve is the Fourier transform
of the simulated muon polarization function which best fits the
data and the shaded region is the residual background signal.}
\end{figure}
In a $d_{x^2-y^2}$-wave superconductor 
the magnetic field distribution
in the core region can be fourfold symmetric \cite{Soininen:94,Ichioka:96}
and twofold symmetric with $a$-$b$ anisotropy \cite{Xu:96}.
However, theoretical models for a $d_{x^2-y^2}$-wave vortex core
contain too many parameters to be useful in fitting experimental
data. Modelling the core region with circular (or elliptical)
symmetry should be sufficient to characterize the changes in core size
with field and temperature. In general one expects
the symmetry of the $d_{x^2-y^2}$-wave vortex cores 
to distort the lattice from triangular symmetry, but only
at high fields where the intervortex spacing is small.
So far no experiments have imaged 
the vortex lattice in YBa$_2$Cu$_3$O$_{6.60}$.

Figure~1 shows the real amplitude of the Fourier transform 
of the muon precession signal in YBa$_2$Cu$_3$O$_{6.60}$ 
for $T \! = \! 0.04~T_c$ and $H \! = \! 0.75$~T (solid curve)
and of the simulated muon polarization function which
best fits the data (dashed curve).
Above $T_c$ the lineshape is nearly a perfect
Gaussian with a width entirely due to the nuclear-dipolar fields.
Well below $T_c$ the internal field distribution is very similar to that
previously observed in NbSe$_2$, where the
vortex lattice is known to be triangular.
A small peak due to a residual background signal is also visible.
At $0.84~T_c$ the field distribution is no longer asymmetric (Fig.~1 inset).
We attribute this qualitative change in the lineshape to melting
of the vortex lattice, ie: a transition from continuous 3D vortex lines 
to 2D 
\begin{figure}[t]
\epsfig{figure=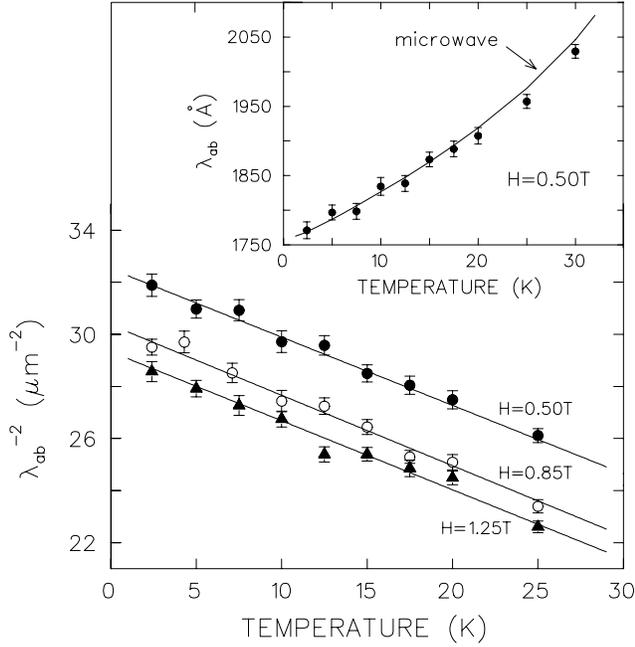, width=\linewidth}
\\
\caption[]{
The temperature dependence of $\lambda_{ab}^{-2} (T,H)$
in YBa$_{2}$Cu$_{3}$O$_{6.60}$ (S2)
for applied fields of 0.5 (solid circles), 0.85 (open circles) 
and 1.25~T (solid triangles).
Inset: The $T$-dependence of $\lambda_{ab}$ at 0.5~T.
The solid line shows the microwave measurements 
of $\Delta \lambda_{ab}(T) \! = \! \lambda_{ab}(T)-\lambda_{ab}(1.25~K)$ 
in zero field \cite{Bonn:96} 
assuming our value $\lambda_{ab}(1.25~K) = 1762$~\AA.}
\end{figure}
vortex ``pancakes'' which are uncorrelated between planes.
Such a transition has been observed in $\mu$SR studies of highly anisotropic
Bi$_{2.15}$Sr$_{1.85}$CaCu$_2$O$_{8+ \delta}$ (BSCCO) \cite{Lee:95}
but does not occur in YBa$_2$Cu$_3$O$_{6.95}$ in similar fields
because of stronger interplane coupling.
We estimate the variation of the crossover temperature $T_m$ with 
magnetic field for $0.5 \! < \! H \! < \! 1.5$~T to be
$T_m \! \cong \! T_c - \alpha H$, where $\alpha = 17(1)$~K/T. 
    
Figure~2 shows the temperature dependence of $\lambda_{ab}^{-2} (T)$
below $T_m$ for three magnetic fields. 
As previously observed in YBa$_2$Cu$_3$O$_{6.95}$ \cite{Sonier:97} and 
La$_{1.85}$Sr$_{0.15}$CuO$_{4}$ \cite{Luke:97}, there is
a strong linear-$T$ dependence at low temperature which is independent of $H$.
The inset of Fig.~2 shows a comparison between the $T$-dependence of
$\lambda_{ab}(T)$ at 0.5~T and microwave cavity measurements \cite{Bonn:96}
of $\Delta \lambda_{ab}(T)$ in zero magnetic field.
The excellent agreement confirms that the fitting procedure
which assumes a triangular vortex lattice,
introduces at most only a small systematic error in the absolute
value of $\lambda$. This is reasonable since it has been shown
theoretically that including additional terms in the free energy
of the vortex state produce only minor changes in the internal
field distribution \cite{Affleck:96}.

As in Ref.~\cite{nbse2:97} we define $\rho_0$ to be the
radius at which the supercurrent density
${\bf J}_s(\mbox{\boldmath $\rho$}) \! = \!
\mbox{\boldmath $\nabla$} \! \times \! {\bf B}(\mbox{\boldmath $\rho$})$
reaches its maximum value. $J_s(\rho)$ profiles were generated from
fits of the data to the field profile
of Eq.~(\ref{eq:Br}). Figure~3 shows the $T$-dependence 
of $\rho_0(T)$ for the same fields as in Fig.~2. The solid lines are
fits to the linear relation
\begin{figure}[t]
\epsfig{figure=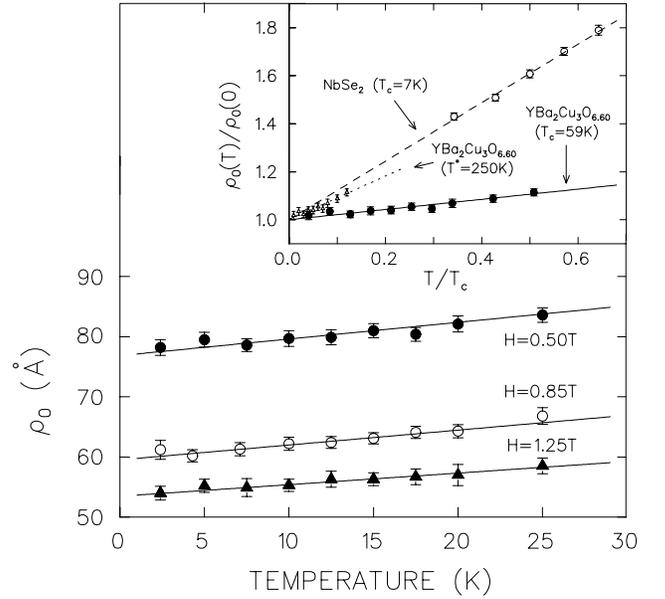, width=\linewidth}
\\
\caption[]{The $T$-dependence of $\rho_0$
in YBa$_{2}$Cu$_{3}$O$_{6.60}$ (S2) 
for applied fields of 0.5 (solid circles), 0.85 (open circles) 
and 1.25~T (solid triangles).
Inset: The $T$-dependence of $\rho_0(T)/\rho_0(0)$ in
NbSe$_2$ at 0.19~T (open circles) and in
YBa$_{2}$Cu$_{3}$O$_{6.60}$ at 0.5~T normalized to $T_c\!=\!59$~K (solid
circles) and $T^*\!=\!250$~K (open triangles).}
\end{figure}
$\rho_0 (T) \! = \! \rho_0(0)[1+\beta T/T_c]$,
where $\beta \! \sim \! 0.23$ is essentially independent of field.
The inset of Fig.~3 shows the $T$-dependence of
$\rho_0$ at 0.5~T normalized to $\rho_0$ at $T\!=\!0$.
The linear term is much weaker than that found in
NbSe$_2$ at 0.19~T \cite{footnote}, where 
$\beta \! \sim \! 1.2$.
Better agreement (see Fig.~3) is obtained if $T$
is normalized to $T^*\!=\!250$~K, the temperature below which
a pseudogap opens in the spectrum of low-energy excitations
in YBa$_2$Cu$_3$O$_{6.60}$ (Ref.~\cite{Basov:96}).
One possible interpretation is that the pairing amplitude 
is established prior to the onset of 
long range phase order at $T_c$ \cite{Emery:95}.
In both materials the size of the vortex core does not decrease
as steeply with temperature as expected in theoretical predictions
for a $s$-wave \cite{Kramer:74} or
$d_{x^2-y^2}$-wave \cite{Ichioka:96} superconductor.
However, these theories pertain to a single isolated
vortex and do not account for vortex lattice effects.
Thermal fluctuations of the vortices about their average
positions result in a 
premature truncation of the
high-field tail in the $\mu$SR lineshape---which results in an 
overestimate of $\rho_0$ that increases with $T$.
However, thermal fluctutations are expected to be most important 
at high magnetic fields \cite{Cubitt:93} or near $T_m$, and do not account for
the weaker $T$-dependence relative to NbSe$_2$.

In Fig.~4, $\lambda_{ab}$, $\rho_0$
and $\kappa$ extrapolated to $T$=0 are plotted as a function of $H$. 
The magnitude of $\lambda_{ab}$ determined in S1 is significantly
lower than that in S2. The difference is likely a result of
vortex lattice distortions in S1 due to twin boundary 
pinning. This introduces a systematic uncertainty in the determination 
of $\lambda_{ab}$. The RMS deviation of the vortices 
from their positions in a perfect 
\begin{figure}[t]
\epsfig{figure=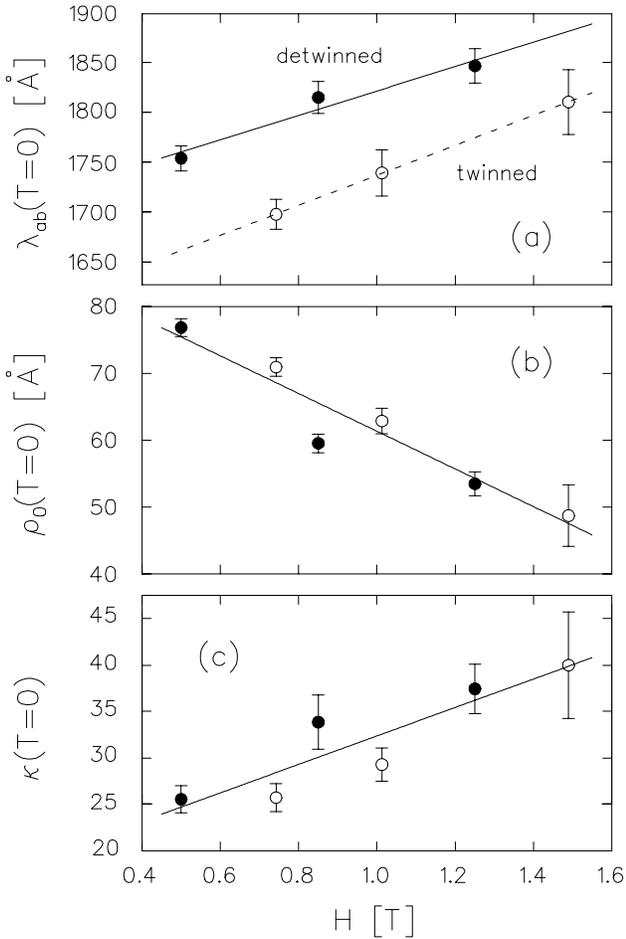, width=\linewidth}
\\
\caption[]{The field dependence of
(a) $\lambda_{ab}$ (b) $\rho_0$ and (c) $\kappa_{ab}=\lambda_{ab}/\xi_{ab}$
extrapolated to $T \! = \! 0$ in 
YBa$_{2}$Cu$_{3}$O$_{6.60}$. The data for S1 (twinned) are shown
as open circles whereas the solid circles designate S2 (detwinned).}
\end{figure}
triangular lattice (estimated from
the fitted values of $\sigma$) was found to be $\sim \! 8$\% and 5\%
of the intervortex spacing for S1 and S2, respectively.
This disorder was independent of $H$.
The lines in Fig.~4(a) are fits to the linear relation,
$\lambda_{ab} (0,H) \! = \! \lambda_{ab}(0,0)[1+\gamma H/H_{c2}]$.
Assuming $H_{c2}\!=\!70$~T, $\lambda_{ab}(0,0) \! = \! 1586$~\AA~and 
$\gamma \! = \! 6.6$ in S1, and
$\lambda_{ab}(0,0) \! = \! 1699$~\AA~and 
$\gamma \! = \! 5.0$ in S2.
The increase in $\lambda_{ab}$ with $H$ is comparable to that
observed recently in YBa$_2$Cu$_3$O$_{6.95}$ \cite{Sonier:97}
and is considerably stronger than that reported in NbSe$_2$
where $\gamma \! = \! 1.6$ \cite{nbse2:97}. This difference 
can be attributed to an enhancement in
pair breaking caused by the applied field for an energy gap 
function with nodes---analogous to the nonlinear Meissner
effect \cite{Yip:92}.

Figure~4(b) shows how $\rho_0$ decreases with increasing magnetic field.
Twin boundaries appear to have a negligible effect on the core
size since there is good agreement between S1 and S2.
A physical interpretation is that the increased interaction 
between vortices with field ``squeezes'' the vortex cores causing
a reduction in $\rho_0$. The microscopic theory for a conventional
superconductor predicts such behaviour 
\cite{Golubov:94}.
It is conceivable, the result may be attributed
to quantum fluctuations at low temperatures which become important
for large $H$ \cite{Blatter:94}. However, quantum fluctuations
are likely to have a small effect given the observed field dependence
of $\rho_0$ in NbSe$_2$ (Refs.~\cite{Hartmann:93,nbse2:97}).
The essential point is that quantum fluctuations are expected to be
negligible in a conventional superconductor where $\xi$ is large
\cite{Blatter:94}. 
It is important to realize that our measurements in
YBa$_2$Cu$_3$O$_{6.60}$
extend over a very narrow range of $H/H_{c2}$ relative to those
for NbSe$_2$. We were limited to this field
range by the melting transition for higher $H$ and the
small amplitude of the high-field cutoff for lower $H$. 
Consequently, no reduction in the rate of
change of $\rho_0$ with $H$ (expected for larger fields) was observed.
A linear fit to the data yields an intercept 
$\rho_0 (0,0) \! = \! 89.5$~\AA~and slope $-28.2$~\AA/T.

Figure~4(c) shows the $H$-dependence of $\kappa \! = \! \lambda_{ab}/\xi_{ab}$.
A linear fit gives an intercept $\kappa (0,0) = 17.0$ and slope
$15.3$~T$^{-1}$. We also find that  
$\kappa$ is essentially independent of $T$ over the entire temperature 
range. This agrees not only 
with magnetization measurements in the vortex state of
BSCCO for $T/T_{c} \! > \! 0.43$ \cite{Schilling:94}, but also
with our previous measurements in NbSe$_2$. 

In conclusion, we have measured the $T$ and $H$-dependences 
of $\lambda_{ab}$ and $\rho_0$ in YBa$_2$Cu$_3$O$_{6.60}$
for $H \! \ll \! H_{c2}$.
We find that $\lambda_{ab}$ and $\rho_0$ both vary linearly with $T$ 
at low temperatures. The $T$-dependence for $\rho_0$
is considerably smaller than that found in NbSe$_2$.
Also, $\lambda_{ab}$ increases
while $\rho_0$ decreases with increasing magnetic field. 
We attribute the field dependence of $\rho_0$
to the compression of the vortices due to
vortex-vortex interactions.

\begin{center} 
\rule{0.3\textwidth}{0.25pt} 
\end{center}  
We would like to thank Alain Yaouanc, Ian Affleck,
John Berlinsky, Catherine Kallin, 
and Marcel Franz for many helpful 
discussions, and Syd Kreitzman, Curtis Ballard 
and Mel Good for technical assistance.  
This work is supported by NSERC of Canada and by the
U.S. Department of Energy through grant
DE-FG05-88ER45353.  
    
 
\end{document}